\documentclass[acmsmall]{acmart}
\usepackage{listings}
\usepackage{xspace}
\usepackage{multirow}

\definecolor{lightgray}{gray}{0.1} % 0.6表示灰度级别

\newcommand{\titleicon}{%
  \ifnum\value{page}=1
    \includegraphics[width=1.0em]{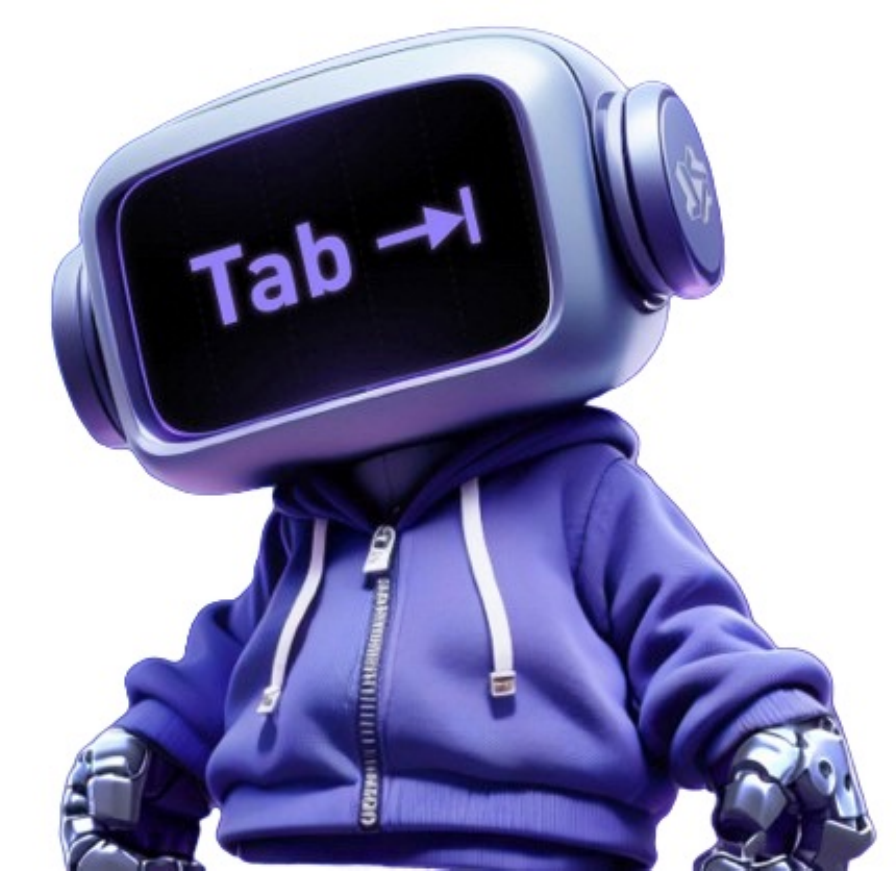}%
  \fi
}

\AtBeginDocument{%
  }
\newcommand{\ourmethod}{EvoCoder\xspace}

\setcopyright{acmlicensed}
\copyrightyear{2018}
\acmYear{2018}
\acmDOI{XXXXXXX.XXXXXXX}

\acmConference[arxiv'25]{arxiv}{June 03--05,
  2018}{Woodstock, NY}
\acmBooktitle{Woodstock '18: ACM Symposium on Neural Gaze Detection, June 03--05, 2018, Woodstock, NY}
\acmISBN{978-1-4503-XXXX-X/18/06}

\usepackage{graphicx} % 用于插入图片
\usepackage{subcaption} % 用于创建子图
\usepackage[most]{tcolorbox}
\usepackage[normalem]{ulem}

\setcopyright{none}
\makeatletter
\renewcommand\@formatdoi[1]{\ignorespaces}
\makeatother
\renewcommand\footnotetextcopyrightpermission[1]{} % removes footnote with conference information in first column
\begin{document}

\title{LLMs as Continuous Learners~\titleicon: Improving the Reproduction of Defective Code in Software Issues}

% \title{LLMs as Continuous Learners: Improving the Reproduction of Defective Code in Software Issues}

%% Of note is the shared affiliation of the first two authors, and the "authornote" and "authornotemark" commands
% \affiliation{%
\author{Yalan Lin\textsuperscript{*}}
\affiliation{%
 \institution{Shanghai Jiao Tong University}
 \country{China}
}
\thanks{\textsuperscript{*} Work during Yalan’s internship at Tongyi Lab@Alibaba group.}

\author{Yingwei Ma, Rongyu Cao, Binhua Li, Fei Huang}
\email{mayingwei.myw@alibaba-inc.com}
\affiliation{%
 \institution{Tongyi Lab, Alibaba Group}
 \country{China}
}
% \author{Rongyu Cao}
% \affiliation{%
%  \institution{Tongyi Lab, Alibaba Group}
% }
% \author{Binhua Li}
% \affiliation{%
%  \institution{Tongyi Lab, Alibaba Group}
% }
% \author{Fei Huang}
% \affiliation{%
%  \institution{Tongyi Lab, Alibaba Group}
% }
\author{Xiaodong Gu\textsuperscript{\textdagger}}
\email{xiaodong.gu@sjtu.edu.cn}
\affiliation{%
 \institution{Shanghai Jiao Tong University}
 \country{China}
}
\author{Yongbin Li\textsuperscript{\textdagger}}
\affiliation{%
 \institution{Tongyi Lab, Alibaba Group}
 \country{China}
}
\thanks{\textsuperscript{\textdagger}Corresponding Authors.}

% \affiliation{%

\renewcommand{\shortauthors}{Trovato et al.}

\begin{abstract}
%Defective code poses a direct threat to the reliability and scalability of software systems. Reparing defective code requires not only the regular identification and correction of newly emerging errors (i.e., bugs) but also the creation of corresponding test functions to validate the effectiveness of these modifications. 
 Reproducing buggy code is the first and crucially important step in issue resolving, as it aids in identifying the underlying problems and validating that generated patches resolve the problem. While numerous approaches have been proposed for this task, they primarily address common, widespread errors and struggle to adapt to unique, evolving errors specific to individual code repositories. To fill this gap, we propose \ourmethod, a multi-agent continuous learning framework for issue code reproduction. \ourmethod adopts a reflection mechanism that allows the LLM to continuously learn from previously resolved problems and dynamically refine its strategies to new emerging challenges. To prevent experience bloating, \ourmethod introduces a novel hierarchical experience pool that enables the model to adaptively update common and repo-specific experiences. Our experimental results show a 20\% improvement in issue reproduction rates over existing SOTA methods. Furthermore, integrating our reproduction mechanism significantly boosts the overall accuracy of the existing issue-resolving pipeline.
\end{abstract}




\received{20 February 2007}
\received[revised]{12 March 2009}
\received[accepted]{5 June 2009}

\maketitle
\section{Introduction}

% 测试代码在软件仓库中扮演着至关重要的角色，确保了软件的质量与稳定性。通过自动化测试，不仅可以及时发现并修复错误，还能提高开发效率，是维护项目长期健康发展的重要基石。然而，观察实际代码仓库可以发现，很多测试代码并非在项目初期就已编写完成，而是在用户报告相关bug后，根据这些bug的具体情况新增的。因此，基于issue描述自动生成复现问题的代码成为了一种有效的测试用例生成方法。这种方法不仅能够加速问题定位与解决过程，还促进了持续集成与交付流程中的质量保证活动，从而增强了软件系统的健壮性和可靠性。

% 本文主要探讨如何根据issue描述生成高质量的复现代码。具体来说，给定一个仓库和对应的issue，该issue可能包含复现问题的核心步骤（如图1.a所示），也可能仅有自然语言描述而无具体步骤（如图1.b）。我们的目标是构建一种模型，能够生成一段可运行的代码，其输出与issue中提到的“实际结果”一致；而在应用正确的补丁修复此bug后，这段代码的输出则应与“预期结果”相匹配。

% 已有研究对复现代码的作用进行了广泛探索，在一些代码相关的智能代理（例如CodeR与SWE-Agent）中已经集成了这一功能。此外，某些工作直接从issue生成测试用例，而在ACR和其他无代理方法中也强调了添加测试用例以过滤和调试生成代码的重要性，这有助于显著提升最终代码的质量。尽管如此，当前解决方案普遍存在准确性较低的问题，错误的复现代码甚至可能会误导其他部分的智能代理，产生负面影响。

% 在本研究中，我们首先进行了实证研究，随后基于实证结果设计了一种持续学习机制，使模型能够从以往解决问题的经验中总结教训，并将这些经验应用于新问题的解决过程中。最后，我们将所提出模型的代码复现效果与其他现有模型进行了对比，并在SWE-bench上评估了该模型在整个代码缺陷修复任务中的表现提升。

% 总之，本文的主要贡献包括：

% + 对先前模型复现失败的原因进行了深入分析，找出了导致复现失败的根本原因。
% + 提出了一种新的优化方法来提高代码复现的效果，使得复现成功率提升了20\%，并且这种方法还具有扩展到代码生成等其他任务的潜力。

Issue resolving is a fundamental aspect of software development and maintenance, critical to preserving the quality and stability of software systems~\cite{zuber1998integrated, zhang2016literature, xia2013accurate, ma2024lingma}. Throughout a project's lifecycle, various issues inevitably arise. Automating the resolving of these issues not only accelerates error identification and correction but also boosts development efficiency, playing a pivotal role in sustaining the long-term health of the project~\cite{jimenez2023swe, xia2024agentless, zhang2024autocoderover, ma2024understand}. This process is especially vital in the early stages of development, where test cases are often incomplete, requiring continuous refinement as new bugs are reported by users~\cite{kang2023large}.

\begin{figure}[htbp]
    \centering
    \includegraphics[width=\linewidth]{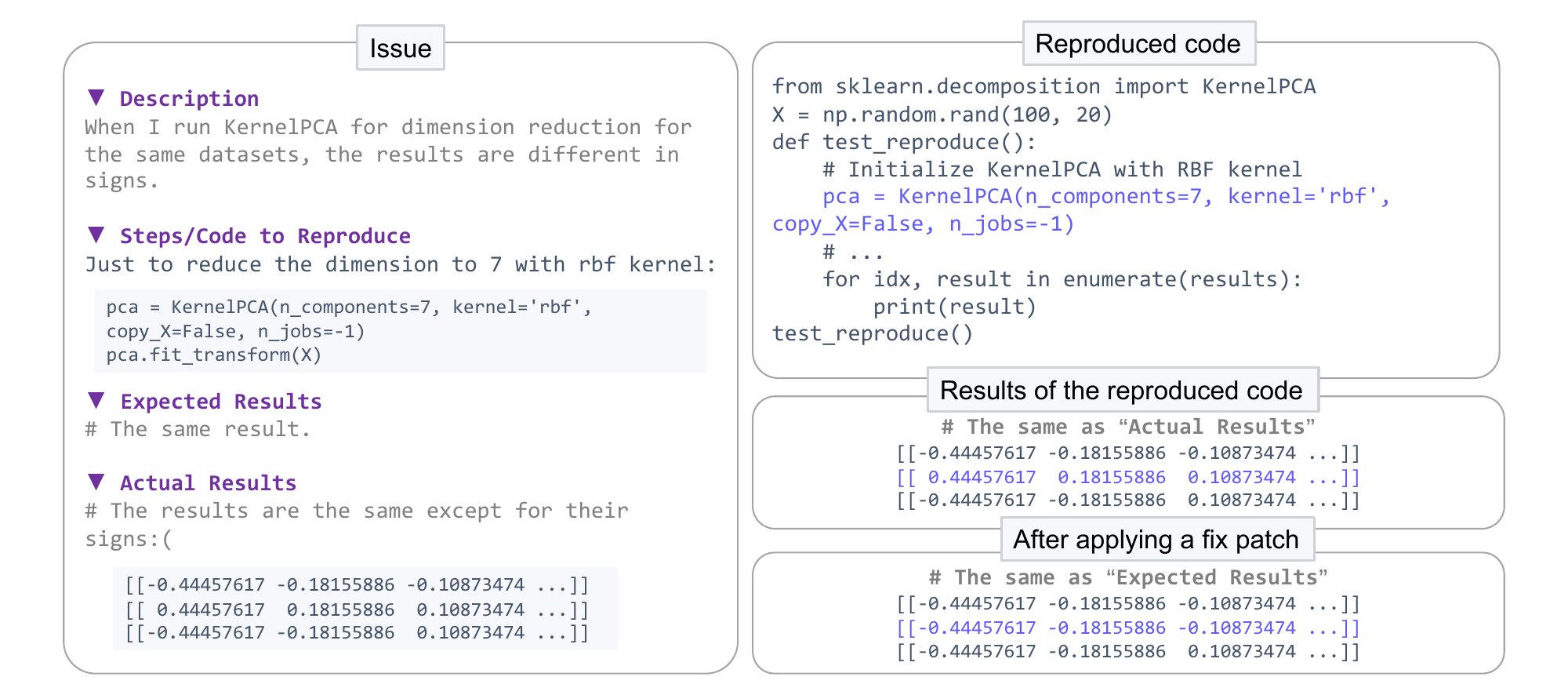}
    \caption{An example of issue reproduction for the KernelPCA library. The \textit{Expected Results} section specifies the ground truth results for the original problem. The \textit{Actual Results} section specifies the erroneous results that need to be reproduced based on the given steps. As seen in the upper right, the reproduced code successfully produces the ``Actual result'' in the issue (the three vectors are different in sign). To resolve this issue, a patch is needed to fix the bug. As seen in the lower right, applying the patch aligns the result with the ``Expected result'', ensuring that all three vectors are now correctly uniform in sign.}
    \label{fig:intro}
\end{figure}

In issue resolving, the initial and critical step is issue reproduction which involves automatically generating code to reproduce the reported problem based on the issue description. 
Specifically, given an issue from a code repository, which contains key steps to reproduce the problem (as shown in Figure \ref{fig:intro}), the goal is to generate executable code to replicate the issue. The output of this generated code should match the ``Actual Result'' specified in the issue while applying an appropriate patch should modify the code’s output to reflect the ``Expected Result''.
Successful issue reproduction not only accelerates the process of problem localization and resolving but also strengthens quality assurance processes in continuous integration and delivery, thereby improving the robustness and reliability of software systems. %On the other side, incorrect reproductions might mislead other parts of the intelligent agents, leading to negative consequences.

Previous studies have extensively explored the role of reproducing code \cite{chen2024coder,yang2024swe,zhang2024autocoderover}. Some research integrates this process into code intelligent agents (e.g., CodeR~\cite{chen2024coder} and SWE-Agent \cite{yang2024swe}). These methodologies design computer interfaces for viewing, searching, and editing files, thereby enabling LMs to resolve issues by chatting with computers as humans do. In contrast, agent-less approaches such as AutoCodeRover \cite{zhang2024autocoderover} and Agentless~\cite{xia2024agentless} generate test cases from issues, and use test cases for filtering and debugging the reproduced code. 

Despite these advancements, current solutions primarily address common, widespread errors and struggle to adapt to unique, evolving errors specific to individual code repositories. For example, some issues depend on project-specific formatting conventions to accurately reproduce. This might involve a pipeline where dependency libraries are introduced first, followed by defining the reproduction function within a \emph{test\_reproduce} function, and subsequently calling this function. Additionally, some issues rely on human-defined rules that LLMs struggle to interpret and apply effectively. These project-specific conventions are typically unique to each repository and, as such, are not easily transferable across different projects.
A more comprehensive illustration of the motivation will be presented in Section~\ref{sec:motivation}.

In this research, we propose \ourmethod, a novel continual learning framework for issue code reproduction. 
\ourmethod introduces a hierarchical reflection architecture where an actor LLM reproduces the issue code and stores the action trajectory in the memory. A Reflection LLM distills experiences from its trajectory. To better maintain and utilize the extracted experience, we design a hierarchical experience pool: the higher layer stores general experiences while the lower layer corresponds to repository-specific experiences. To keep the experiences up-to-date and continuously refined, we define five actions including \textsc{Add}, \textsc{Remove}, \textsc{Merge}, \textsc{Endorse}, and \textsc{Modify}, for the reflection LM to manipulate the hierarchy of experiences. 
Our approach enables the model to continuously learn and optimize as it encounters new problems, without incurring additional computational costs associated with fine-tuning. 
In addition, it dynamically updates and refines the model's existing knowledge base, allowing it to develop expertise within a specific domain.
%\ourmethod allows the model to learn from past experiences in resolving issues and apply these lessons to new issues. 

We evaluate the effectiveness of the \ourmethod on SWE-bench \cite{jimenez2023swe} and compare it with the state-of-the-art issue code reproduction methods. Experimental results reveal that \ourmethod can improve the reproduction rate by 20\%. In addition, the generated reproduction code successfully assists in the model's debugging process, thereby enhancing the accuracy of issue resolving.

The main contributions of our paper are threefold:
\begin{itemize}
\item An in-depth analysis of the root causes behind the failure of previous models on issue reproduction.
\item We propose a novel issue reproduction method based on continual learning and reflection LM, resulting in a 20\% increase in success rate.
\item Extensive evaluation of the proposed method on state-of-the-art issue resolving benchmarks.
\end{itemize}

\section{Background}

\subsection{Issue Resolving}

Issue resolving is a critical component of software development and maintenance, encompassing the identification, diagnosis, and resolving of software defects or issues reported by users or developers. In this task, based on the provided problem descriptions and snapshots of the codebase, the model needs to autonomously identify the specific locations that need modification~\cite{zhang2024autocoderover, wang2024codeact, ma2023mulcs, huang2023towards}. This can be achieved through commands to search for certain files or by analyzing execution paths using existing passing and failing test cases to pinpoint the most likely points of failure. Once these locations are identified, the model generates the corresponding code snippets to address the issues~\cite{yu2023wavecoder, luo2023wizardcoder, wei2024magicoder, zhu2024deepseekcoder}.
During the generation process, the model can also perform debugging and iterative revisions on its own~\cite{chen2023teaching, ding2024cycle, shi2024code, ni2024next, yan2024better}. For instance, syntax checkers can identify syntactical errors, and test cases or code snippets that reproduce the issue can be executed to verify whether the generated patches resolve the original errors. Any error messages or feedback from these processes can be fed back into the model to further refine and improve the generated code.
In the final testing phase, the generated code patches must pass both the unit tests associated with the current modifications—extracted from the pull requests submitted by programmers—and the existing unit tests. This ensures that the new changes do not adversely affect previously implemented functionalities, the generated code patches must pass both the unit tests associated with the current modifications—extracted from the pull requests submitted by programmers—and the existing unit tests. This ensures that the new changes do not adversely affect previously implemented functionalities.

\subsection{Issue Code Reproduction}
Issue Reproduction involves the creation of executable code designed to reproduce the issues reported by users. Historically, this challenge has been framed as a single-step code generation task, wherein the model processes the issue description provided by the user and outputs the corresponding code snippet intended to recreate the reported problem~\cite{kang2023large}. However, with advancements in LLMs, there has been a shift towards models capable of engaging in multi-turn dialogues. This evolution allows the models to not only generate initial code but also to interact with a simulated environment for debugging purposes. Prominent instances of this advanced approach can be found in systems like SWE-Agent~\cite{yang2024swe} and CodeR~\cite{chen2024coder}, both of which utilize agent-based methodologies. In these frameworks, the interaction begins with the model receiving a system prompt that outlines the range of actions it can undertake, including edit, search, and other relevant tasks. Subsequently, the user inputs the specifics of the issue they wish to address. Operating in a ReAct paradigm~\cite{yao2022react}, the model proceeds to analyze the current state, contemplate potential steps, and execute the most suitable action. This typically involves generating the necessary files, modifying them to integrate the required features, and systematically debugging the program until it faithfully replicates the 'actual result' specified in the original issue report, or halting if multiple attempts fail to yield further progress.

In our work, we continue to build upon this agent-based paradigm, exploring how models can achieve sustained improvement in the process of reproduce a wide variety of issues, thereby refining its ability to accurately and efficiently reproduce issues over time.

\section{Motivation}
\label{sec:motivation}

% \lin{TODO: 给一个codeR复现过程的示意图。 地下直接拆分两部分，提取和判断标准都分开讲一下。}

% 为分析之前的方法复现失败的原因，我们从CodeR的history中抽取了其生成的复现代码，并运行代码观察输出，人工判断了一段代码是否正确，判断标准如下：
% 1. 若题目中提供了核心复现代码或完整代码，则复现代码必须严格遵循所提供内容，严禁遗漏任何关键语句。
% 2. 若题目中包含运行指令，则这些指令应与题目描述完全一致。
% 3. 若题目中提供了完整的报错信息，则运行时的报错应保持一致；若仅有自然语言描述，则输出的信息应能有效反映该问题。
% 4. 若问题不涉及缺陷修复而是新功能添加，则输出应能准确反映调用该功能后的结果。
% 5. 报错信息应源于实际运行结果，而非通过输出等方式进行模拟。

In this section, we motivate our approach by analyzing the failure cases of CodeR \cite{chen2024coder}, the state-of-the-art method for issue reproduction. We run CodeR on SWE-bench-lite \cite{swebenchlite} and extract reproduction code generated by CodeR from its resolving trajectories. Upon running the generated issue code, we manually inspect the execution output and assess its correctness based on the following criteria:

\begin{enumerate}
\item \textbf{Completeness}: The reproduced code must contain the core or complete code provided by the users in the original issue.
\item \textbf{Consistency}:
(i) The error messages upon running the code should be consistent with the provided full error messages.  If the input issue involves adding a new feature, the output should contain the new feature.
(ii) Execution commands involved in the code must match those described in the problem statement exactly.
\item \textbf{Authenticity}: Error messages must be derived from actual runtime results, instead of mocked outputs.
\end{enumerate}

Ultimately, we collect a total of 84 erroneous codes generated by CodeR. Through manual examination, we identified seven types of errors that can be categorized into two groups: 

\textbf{Internal errors (58.55\%)}
refers to intrinsic errors that arise from the reproduction code itself %In the entire process of reproduction code generation, i.e., from identifying the reproduction goal to reproducing the errors, and ultimately displaying the error message in the issue, there can be the following five types of errors. 
, including
1) \textit{Wrong Reproduction Targets (4.00\%):} The execution results from the reproduced code do not align with the expected error message described in the original issue. This is likely due to the model's difficulty in fully interpreting the natural language within the issue description—such as distinguishing between error outputs and correct outputs.
2) \textit{Wrong Function Call (10.97\%):} the reproduced code calls wrong functions or commands compared to the intended issue code.
3) \textit{Over-mocking (14.63\%):} Mocking is a commonly used technique in software testing that isolates test scripts from external dependencies by replacing them with mocked functions. However, when the behavior of these dependencies is overly predefined, the mocked functions may not accurately reflect the actual behavior of the original code. For instance, using print statements to output error messages instead of triggering real errors can cause tests to pass even if underlying defects exist. This limits the tests’ effectiveness in detecting issues, especially when the true behavior of dependencies changes.
4) \textit{Missing Environment Requirements (6.10\%):} The reproduction only contains the core code while missing the environmental setup. 
5) \textit{Inaccurate Execution Results (23.17\%):} The reproduced code contains logical errors during the reproduction process, causing inconsistent results to those described in the issue. This is probably due to the restriction of model capability and the task complexity. %\gu{what is the difference between 1) and 5)?}\lin{1) is the target is wrong(e.g. I need to check the content of error message, but it only output "error". 5) is it try to reproduce it, but the logic may still have some problem, but is more close to the correct code.}

\textbf{External errors (41.45\%)} refers to the errors from external environmnents or executions, including
1) \textit{Incorrect commands to run the code (14.63\%)}:
Some issues require specific commands for reproduction, which are usually mentioned in the issue description or pertain to particular usages of the repository. However, the previous methods uniformly used Python commands to run the code, which could result in the inability to reproduce certain problems
and 2) \textit{Wrong environment setting (26.82\%)}:
Given that reproducing these issues often requires specific versions of libraries, particular operating systems, and interactive environments (such as Jupyter), it may happen that even with correct code and execution instructions, the reported problem cannot be reproduced if the exact environment is not set up. 

% 仔细观察这些代码复现遇到的错误，我们发现这些错误是存在共性的。有些错误原因在所有库中都普遍存在，比如使用模拟的方式输出错误信息而没有真正复现问题。而有些错误则在特定库中反复出现，例如在Django库中的初始环节设置问题。这使我们联想到人类程序员在不断使用一个库和不断解决遇到的问题的过程中，逐渐熟悉该库的使用并积累相关经验以指导后续的编码。

% 因此，我们为模型设计了一种持续学习的策略，希望通过这种策略使模型能够从每次复现尝试中学习并改进。具体来说，我们引入了一个反思模块，模型在每次复现完成后，总结成功或是失败的经验，并将这些信息用于更新其内部的经验知识库。随着时间的推移，模型将逐步积累更多关于不同库及其常见问题的经验，从而提高其复现代码的准确性和可靠性。通过这种持续学习的策略，我们期望模型能够在长期使用中不断提升其复现能力，最终达到更高的自动化水平和更广泛的适用范围。

While the external errors can be easily handled (e.g., by wrapping the reproduced code with a shell script, including the required environmental modifications and commands for execution),
our focus shifts to the more complex internal errors—often the most challenging aspect of issue reproduction.
Through a comprehensive analysis of errors encountered during code reproduction, we identified two recurring patterns:

\begin{figure}[htbp]
\centering
\begin{subfigure}[b]{\textwidth} % Adjusted width to fit two subfigures side by side
\includegraphics[width=\linewidth, trim=0 40 0 0 clip]{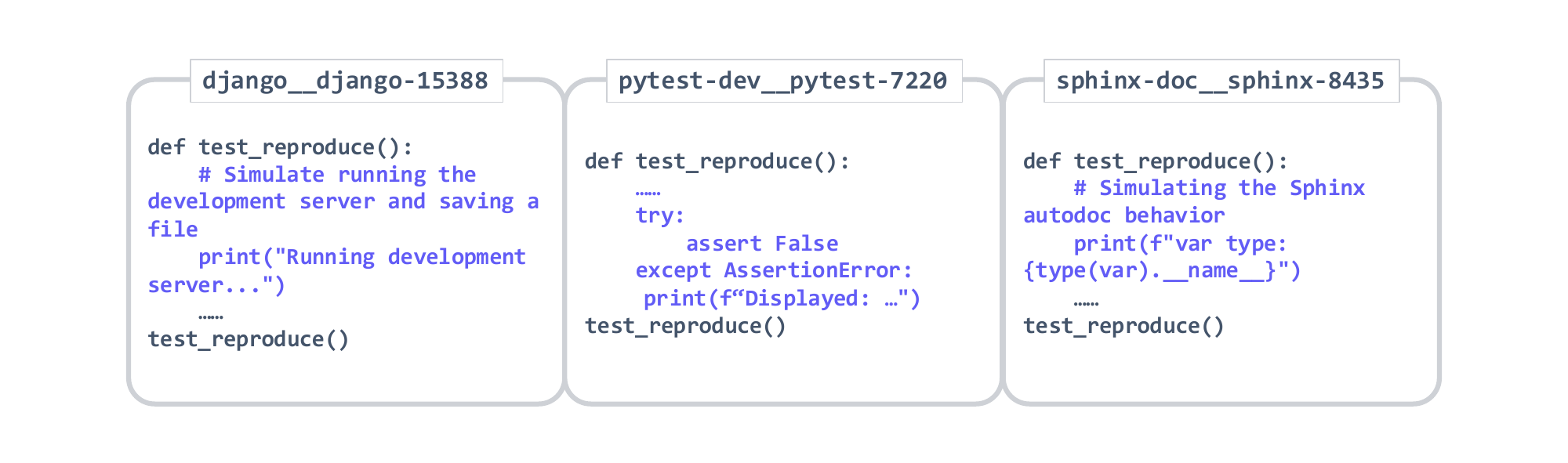}
\caption{Example of the common errors across different libraries.}
\label{fig:case-emp1}
\end{subfigure}
\hfill % Adds horizontal space between subfigures
\begin{subfigure}[b]{\textwidth} % Adjusted width to fit two subfigures side by side
\includegraphics[width=\linewidth, trim=0 40 0 0 clip]{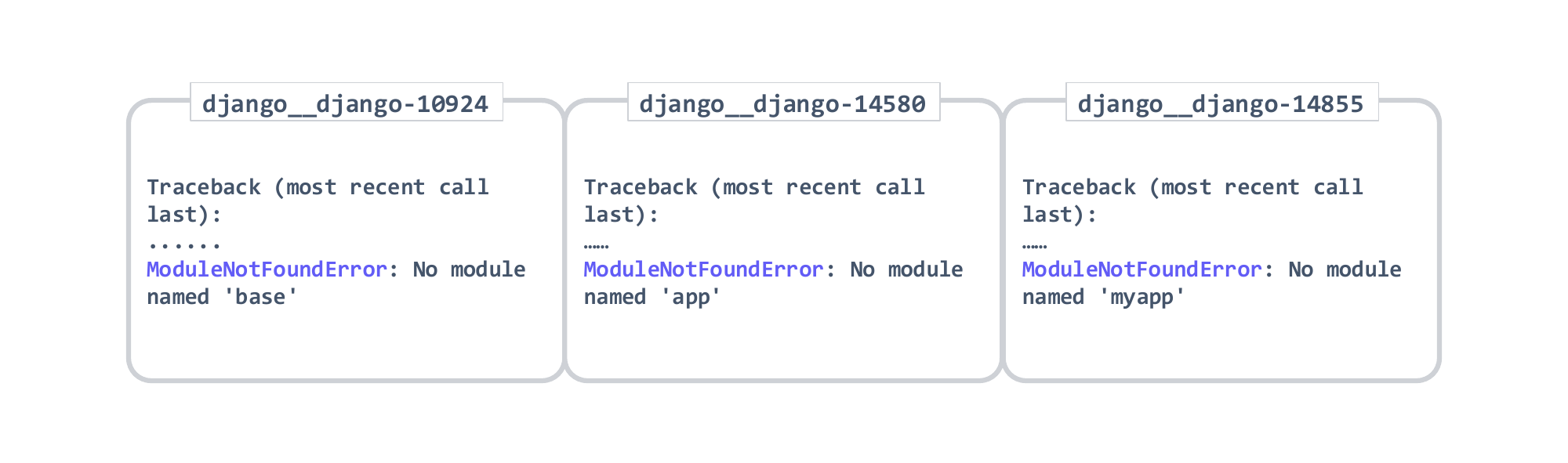}
\caption{Example of recurring errors within the same library.}
\label{fig:case-emp2}
\end{subfigure}
\caption{Examples of the characteristics of the encountered errors.}
\label{fig:motivation_cases}
\end{figure}

\begin{tcolorbox}[width=\linewidth, boxrule=0pt, sharp corners=all,
 left=2pt, right=2pt, top=2pt, bottom=2pt, colback=gray!20]
\textbf{Finding 1}: %The mistakes made by the previous methods stem from two groups: 
Some errors are common across libraries, such as cases that output simulated error messages without reproducing the actual issue (Figure \ref{fig:case-emp1}). Resolving these requires the model to continuously distill general rules from prior resolving experiences.
\end{tcolorbox}

\begin{tcolorbox}[width=\linewidth, boxrule=0pt, sharp corners=all,
 left=2pt, right=2pt, top=2pt, bottom=2pt, colback=gray!20]
\textbf{Finding 2}: Some errors are library-specific, recurring within particular libraries. For example, the errors illustrated in Figure \ref{fig:case-emp2}, which are frequently occurred, are often caused by the initial setup in the Django library. Resolving these issues requires the LLM to adaptively acquire specialized knowledge for each repository.
\end{tcolorbox}

These insights are akin to how human programmers gradually become proficient with a library by repeatedly using the library, solving its issues, and accumulating generic knowledge across different libraries. 
This motivates us with a continuous learning strategy for issue reproduction, where each reproduction attempt accumulates experience, fostering higher levels of automation and broader applicability in issue reproduction.%\gu{the last phrase is not informative}\lin{change to :This motivates us to adopt a continuous learning strategy, where the model reflects and abstracts the experience into memory whenever it finishes reproducing an issue, and utilizes these experience in subsequent reproductions to achieve better results.}

\section{Metohod}

\subsection{Overview}

In this paper, we propose a continuous learning pipeline that enables LLM agents to accumulate experience from previously encountered issues. Unlike conventional methods, our approach does not require fine-tuning; instead, it facilitates the model's ability to continuously update and optimize its stored experiences when addressing new challenges. The overall methodology is structured into three main components:
(i) an \textit{Actor LM} which reproduces the issue using instructions and previous experiences; (ii) a \textit{Reflection LM} which extracts experience from the Actor's reproduction trajectories; and (iii) a hierarchical \textit{Experience Pool} which stores general and repository-specific experiences, enabling the Reflection LM to continuously update and refine its accumulated knowledge.

\begin{figure}[htbp]
    \centering
    \includegraphics[width=\linewidth]{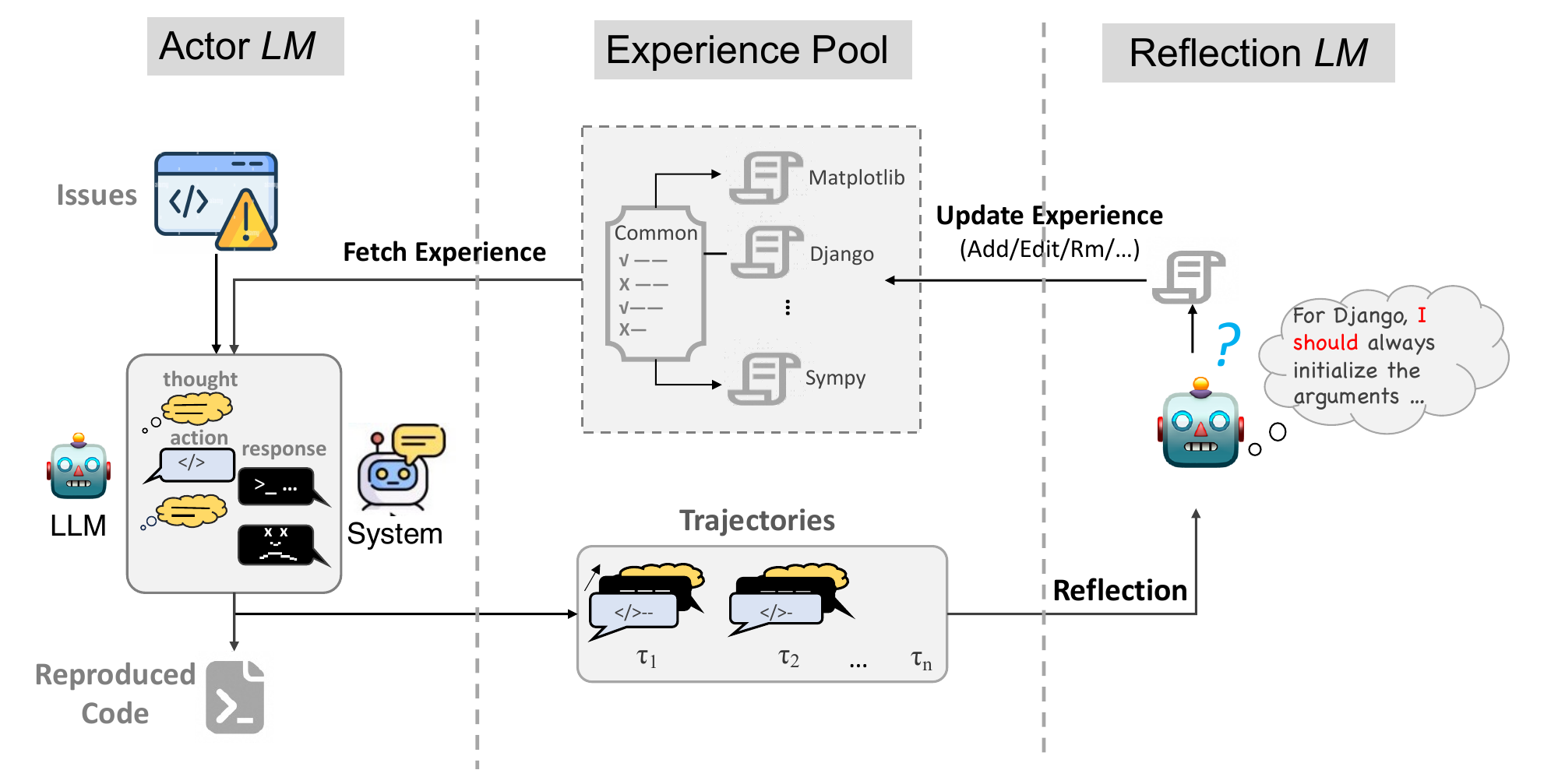}
    \caption{The overall framework of our approach}
    \label{fig:framework}
\end{figure}

\subsection{Actor: Reproduction Trial}

% 我们沿用了SWE-agent，CodeR当中的基本策略来完成代码复现，即把复现过程设计为一个多轮交互的过程，同时为模型定义了查找、查看、编辑文件等多种action，模型每次执行的时候需要输出Thoutht和一个Action，其中的action会被执行，而结果也会返回给模型作为下一次交互的输入。
The Actor is an LLM prompted to perform issue reproduction. 
Adopted from SWE-agent \cite{yang2024swe} and CodeR~\cite{chen2024coder}, we formulate issue reproduction as a multi-turn conversation between agent and computer. In each turn, the Actor is presented with a task instruction, the current system's output, and experiences from past resolving. It is asked to output a ``Thought'' about the current reproducing step, followed by an ``Action'' like searching, viewing, and editing files. Once the reproduction ``Action'' has been executed, the system responds with the new output, which is fed back to the Actor as input for the next interaction. The dialogue continues, yielding a trajectory of Thought-Action-Response sequences.
To mitigate problems caused by improper environment configuration and instruction misunderstanding, we also introduce a standardized process, which asks the model to write instructions to run the code or install related packages in a sh script.
The detailed prompt for the Actor is as follows:

\begin{tcolorbox}[breakable, title=Prompt for Actor, colback=gray!10]
 \scriptsize
  \texttt{We're currently solving the following issue within our repository. Here's the issue text:}\\
ISSUE:
<issue>\\

INSTRUCTIONS:\\
\texttt{You are now going to reproduce the provided issue (not solve the issue). Begin your terminal session in the root directory of the repository. To assist you, use any bash commands or the special interface. Make sure to open and read all the files you need, and execute any tests necessary to reproduce the issue.}\\
\texttt{Remember, YOU SHOULD ENTER ONE COMMAND AT A TIME. Always wait for a response after each command before proceeding.}\\
\texttt{Once you have successfully reproduced the issue and are ready to report it, you can record the steps you took. However, note that you cannot use any interactive session commands (e.g. python, vim) in this environment, but you can run scripts. For example, you can execute a python script with python <script\_name>.py. But DO NOT run './run\_reproduce.sh' or 'bash ./run\_reproduce.sh', use run\_reproduce\_code instead.}\\
\texttt{Once the reproduction is complete, please output the address of the file containing the reproduction script in the following format.}\\

\texttt{NOTE ABOUT THE EDIT COMMAND: Indentation really matters! When editing a file, make sure to insert appropriate indentation before each line! You should also check which file you opened in the editor before editing it.}\\

IMPORTANT TIPS:\\
\texttt{1. Always start by trying to replicate the bug that the issues discusses.}\\
\texttt{If the issue includes code for reproducing the bug, we recommend that you re-implement that in your environment. You MUST create files and execute follow the description, and run it to make sure you can reproduce the bug.}\\
\texttt{If the bug reproduction script does not print anything when it successfully runs, we recommend adding a print("Script completed successfully, no errors.") command at the end of the file, so that you can be sure that the script indeed ran fine all the way through.}\\

\texttt{2. When reproducing the code, you should consider all the cases mentioned in the issue.\\
Before returning, check whether your reproduction of test is complete. The test should be consistent with the issue description. Do not miss the content to be tested. NOTE that the provided code in the issue description of test may be PARTIAL, please generate a complete version of the test code based on the description.}\\

\texttt{3. If the bug reproduction script requires inputting/reading a specific file, such as buggy-input.png, and you'd like to understand how to input that file, conduct a search in the existing repo code, to see whether someone else has already done that. Do this by running the command: find\_file "buggy-input.png" If that doesn't work, use the linux 'find' command.}\\

\texttt{4. If you are uncertain about the specific line number when searching for a snippet of code, a class, or a method within a file, prioritize using the \`grep -nr <code/class/method>\` command to retrieve the approximate location. Then, after opening the file, use the \`goto\` command to navigate to the snippet, preventing excessive use of the \`scroll down\` command. If the \`grep -nr\` command does not retrieve any relevant content, consider using the \`scroll down\` or \`scroll up\` commands to search for the code after opening the file.}\\

\texttt{5. During the reproduction process, if you cannot reproduce due to missing packages in the environment, you MUST use commands like pip, apt-get -y, etc. to install the corresponding packages, please assign exact version of packages if you know when you use pip. Please use --quiet to suppress the output when installing packages. DO NOT install <repo> package as it is already installed and fixed. Remember to write these commands into run\_reproduce.sh.}\\

\texttt{6. You MUST create a file 'run\_reproduce.sh' under the ROOT path of this repo and write the commands into it, including packages you need to install. If you don't need them, simply write commands for code execution into it. Then you can use 'run\_reproduce\_code' to run the code.}\\

\texttt{7. You must execute and modify the file UNTIL you can reproduce the issue. You can create a reproduce.py script for this purpose. If you are able to run the commands directly within a shell script or create other files follow the issue description, then this script is NOT necessary. The structure of your reproduce.py code should output in the following format:}\\

\verb|```| \\
import ...\\
\# setup test environment here\\

\# here is core test code, you MUST use 'test\_reproduce' as function name.\\
def test\_reproduce():\\
\quad \quad  <core code>...\\

\# you MUST call 'test\_reproduce' here.\\
test\_reproduce() \\

\verb|```|  \\

\texttt{8. After editing the files, use 'run\_reproduce\_code' to check whether the issue is reproduced. Please use 'open <file\_name>' to make sure the modified reproduce file have correct logic and STILL satisfies the above format requirements.}\\

\texttt{9. If this issue pertains to the addition of a new feature, your reproduction code should serve to test whether this functionality has been implemented.}\\

\texttt{10. You MUST show the output to the console. If you really need to output the result to a file, you MUST use the 'cat' command to output the result to the console.}\\

\texttt{11. This is the real environment. Please DO NOT use mock or simulation methods to solve the issue. You need to reproduce the issue exactly as described in the issue.}\\

\texttt{12. If you think you have already reproduced the issue, you MUST use 'check' to check whether the issue is reproduced correctly before you submit the code.}\\

\texttt{13. Only return "submit" command when the current task is truly completed. If the current task still cannot be completed after trying many different solutions, please return "fail" command.}\\

\texttt{13. Here are some experiences summarized from other issues in the same repository. Please refer to these experiences during generation to avoid making the same mistakes.}\\

<experience>\\

\texttt{You MUST take into account all the cases mentioned in the issue and reproduce them.}\\

\end{tcolorbox}

%当一个issue解决完之后，会留下一个对话历史文件。这些对话历史不仅记录了解决问题的具体步骤，还隐含着解决问题的关键思路与策略，为进一步优化模型提供了宝贵的数据资源。。一方面，有一些issue当中会给了较为完整的复现代码，其中有一些是该库的独有用法，让模型自己写的话可能写不出来，另一方面，模型的多轮对话中会有一些debug过程，也会对模型如何在一开始就避开某些错误，或是不要尝试一些无用的debug方向起到指导作用。
Upon reproducing the entire issue, the dialogue history, known as the resolving trajectory, is stored in a short-term memory.
The trajectory contains concrete steps, thoughts, and strategies for resolving issues, providing valuable guidelines for issue resolving. For example, some issues could provide complete reproduction code, including specific usages of the library that the model might not be able to generate on its own. The model undergoes debugging steps during the multi-round dialogues, which can guide it in avoiding initial errors or steering clear of unproductive debugging directions.

\subsection{Reflection LM: Extracting Experience from Reproduction Trajectories}

Having collected the trajectories, we distill them into experiences, guiding further issue reproduction. 
An \textit{experience} is defined as a rule guiding what the Agent must do to avoid a certain reproduction failure or follow unique coding styles that certain code repositories may have.  
%One naive idea is directly using these historical dialogues as few-shot examples for the model. However, this method often results in excessively long inputs. Additionally, determining how to recall relevant examples through appropriate similarity measures is a challenging problem. Finally, due to the limitations of the model's capabilities, if the model is not explicitly instructed to think and reason based on these historical records, it is difficult to leverage these examples to aid in the logical generation of subsequent code.
While previous work \cite{liu2024branch,wang2024memory} proposed using experience pools for knowledge accumulation, their approach faces two critical challenges: (1) experience bloat, where experiences become increasingly verbose and detailed over time, hindering accurate issue pattern matching and experience utlizling, and (2) experience rigidity, where experiences become static and fail to adapt to emerging issue types and repository-specific error patterns. 

To enable effective experience refinement, we design an LLM-based reflection mechanism that analyzes task trajectories and golden test patches to continuously update the experience pool.
%Our work addresses these limitations through a novel LLM-based reflection mechanism that not only maintains experience quality but also ensures their adaptability. 
%The reflection mechanism employs five fundamental operations to maintain and optimize the experience pools:, we design five types of Actions for the experience pool: Add\ma{detail?}, Modify, Delete, Approve, and Merge. By employing these Actions\gu{How to manipulate these actions?}
Unlike previous approaches that simply accumulate experiences, our method actively manages experience quality through carefully designed prompts that instruct the reflection LLM to: (1) Analyze Trial. Examines the reproduction trajectory and compares it with the golden test patch to identify successful patterns or failure causes. This analysis determines whether to: \textsc{Add} new experiences when discovering novel reproduction patterns; \textsc{Modify} existing experiences when current solutions can be improved; \textsc{Remove} experiences that consistently lead to failed reproductions. (2) Analyze Rule Applicability. Evaluate whether the experience should be: \textsc{Agreed upon} and kept if it proves effective across multiple repositories; \textsc{Merged} with similar experiences to maintain conciseness; Categorized as repository-specific or generally applicable.

For example, when analyzing a successful reproduction trial, the LLM first compares the issue description with the golden test patch to understand the core reproduction pattern. If this pattern represents a novel approach, it triggers an Add operation; if it refines an existing pattern, it initiates a Modify operation. Through these reflection-based operations, the model continuously improves its reproduction capability in two ways: First, by maintaining a curated set of high-quality experiences that accurately capture issue reproduction patterns; Second, by organizing these experiences into appropriate scopes (general or repository-specific) to facilitate efficient experience utilization.
This ensures that only high-quality experiences are retained and that the total number of experiences remains within a manageable range, preventing it from growing indefinitely with the increasing number of issues.

%然而, 如果直接将历史作为few-shot例子放入模型，可能会非常长，如何通过合适的相似度衡量方式召回到合适的例子也是一个较难的问题，最后受到模型能力限制，如果没有显示让模型依据这些历史做思考和推理的话，很难直接用上这些例子对后续生成代码的逻辑提供帮助。因此，我们在这里用大模型对整个历史对话记录做一个总结，使用如下的prompt，用这种方式把历史记录中后续问题可用的经验浓缩到几句自然语言中，供后续问题参考使用。

\begin{tcolorbox}[breakable, title=Prompt for Reflection LM: Summarize experience from the new trajectory]
\scriptsize
  \texttt{As a software engineering expert, you will be given an issue and summarize experiences from the resolving trajectories. Experiences usually refer to the reason for the reproduction failure and some insights. Pay special attention to this step when the model determines that the reproduction task has failed, but do not follow its format. }\\

  Repository: <repo> \\
  Issue: <issue> \\
  Trajectory: <trajectory> \\

  \texttt{When summarizing your experiences, please carefully compare the issue with the Golden Test Patch. You MUST NOT allow the newly added patch to pass the tests of your reproduction code but fail the Golden Test Patch tests, so please read carefully and summarize which parts are missing from your reproduction code. }\\

  \texttt{Your output should follow the format below. Please only output the list, do not output any other text. }\\

  \texttt{For all repositories: }\\
1. ... \\
2. ... \\

  \texttt{For <repo>:} \\
1. ... \\
2. ... \\

\texttt{If you believe this experience is relevant to all repositories and not just limited to <repo>, please write them after 'For all repositories:'.
If this experience is only applicable to the <repo> repository, write it after 'For <repo>:'.
Note that the content in the two parts should NOT have any repetitions. }
\end{tcolorbox}

\subsection{Hierarchical Experience Pool}
% 为了更好的更新和维护经验，我们设计了总-分的经验池结构，同时使用了多种action的方式来控制经验的数量和质量。
% 结合之前在实证研究中的结论，经验可以分为两类，一类是针对所有仓库的复现过程的general的经验，一类是针对单个仓库独有的经验。为了更好的让模型区分开来这两类经验，我们将经验池设计为总分结构，general的经验是所有issue都可以看到并且共同维护的，special的经验则只有与这个库相关的问题可以看到并且进行维护，《这里需要一个补充》
Based on the findings in Section \ref{sec:motivation}, some error patterns are consistent across repositories, while others are repository-specific. 
To better maintain and utilize the extracted experiences, we design a hierarchical experience pool structure: a general pool at the top level captures common experiences shared across all repositories, followed by a range of repository-specific pools that maintain experiences unique to individual repositories. %\gu{Simply using two separate pools can also achieve this goal}\lin{It seems that this is a general-to-specific structure, where there is a general pool shared by all repositories, followed by individual pools for each repository. Therefore, there should be more than just two pools.},
The common experiences are visible and collaboratively maintained by all issues, whereas repository-specific experiences are only visible and maintained by issues related to the specific repository. This architecture enables the model to distinguish between common and repo-specific experiences more effectively, ensures that different types of experiences are managed and utilized effectively, thereby enhancing the overall performance and adaptability of the model.
% and employed multiple Action types to control the quantity and quality of the experiences.

% 此外，若是直接对经验进行更新，可能会出现经验过长\cite{}, 经验僵化\cite{}等问题，因此参照论文\cite{}的做法，我们将对经验池的操作设计为五种Action，分别为添加、修改、删除、赞同、合并操作，用这种方式让模型在不断解决新issue的过程中，还能对之前总结出的经验进行修正和更新，，确保留下的都是高质量经验，且控制了经验的总条数在一定的范围内不会随着issue个数的增加而无限增长。

\begin{tcolorbox}[breakable, title=Prompt for Reflection LM: updating experience using multiple actions]
\scriptsize
 \texttt{You are an advanced reasoning agent capable of modifying your existing experiences (represented as rules) by adding, editing, removing, or merging rules based on new rules provided. Your task is to update the 'Existing rules' according to the 'New rules' and ensure the final output includes a maximum of 4 operations.}\\

  New rules: \\
  <new\_rules> \\

  Existing rules:  \\
  <existing\_rules> \\

  \texttt{You may perform the following operations:} \\
  \texttt{AGREE: Choose this option if the new rules are present in the existing rules and you think they are very important.}\\
  \texttt{REMOVE: Select this if the new rule contradicts existing rules or if there’s redundancy among the existing rules.}\\
  \texttt{ADD: Use this to introduce new rules that substantially differ from existing ones and are applicable to relevant tasks.} \\
  \texttt{EDIT: Opt for this if an existing rule lacks clarity or generality. Revise it for improvement or to address past issues.} \\
  \texttt{MERGE: Use this to consolidate two similar existing rules into a single, cohesive rule. }\\

  \texttt{Each operation must closely follow the specified format: }\\
  <OPERATION> <RULE NUMBER>: <RULE> \\

  \texttt{The format for each operation is as follows:} \\
  \texttt{AGREE <EXISTING RULE NUMBER>: <EXISTING RULE> }\\
  \texttt{REMOVE <EXISTING RULE NUMBER>: <EXISTING RULE> }\\
  \texttt{EDIT <EXISTING RULE NUMBER>: <NEW MODIFIED RULE> }\\
  \texttt{ADD <NEW RULE NUMBER>: <NEW RULE> }\\
  \texttt{MERGE <EXISTING RULE NUMBER1> <EXISTING RULE NUMBER2>: <NEW RULE> }\\

  \texttt{Please follow the output format: }\\

  \texttt{For all repositories: }\\
  \texttt{1. ADD or EDIT or REMOVE or AGREE or MERGE ... }\\
  2. ... \\

  For <repo>: \\
  \texttt{1. ADD or EDIT or REMOVE or AGREE or MERGE ... }\\
  2. ... \\

  \texttt{Please ensure: }\\
  \texttt{1. There are no repetitions between the "For all repositories" and "For <repo>" sections. }\\
  \texttt{2. If the length of EXISTING RULES is greater than 20, you must use remove or merge at least once. }\\

  \texttt{Below are the operations you do to the above list of EXISTING RULES:} 

\end{tcolorbox}

\subsection{The Reproduction Process}

\begin{figure}[htbp]
    \centering
    \includegraphics[width=\linewidth]{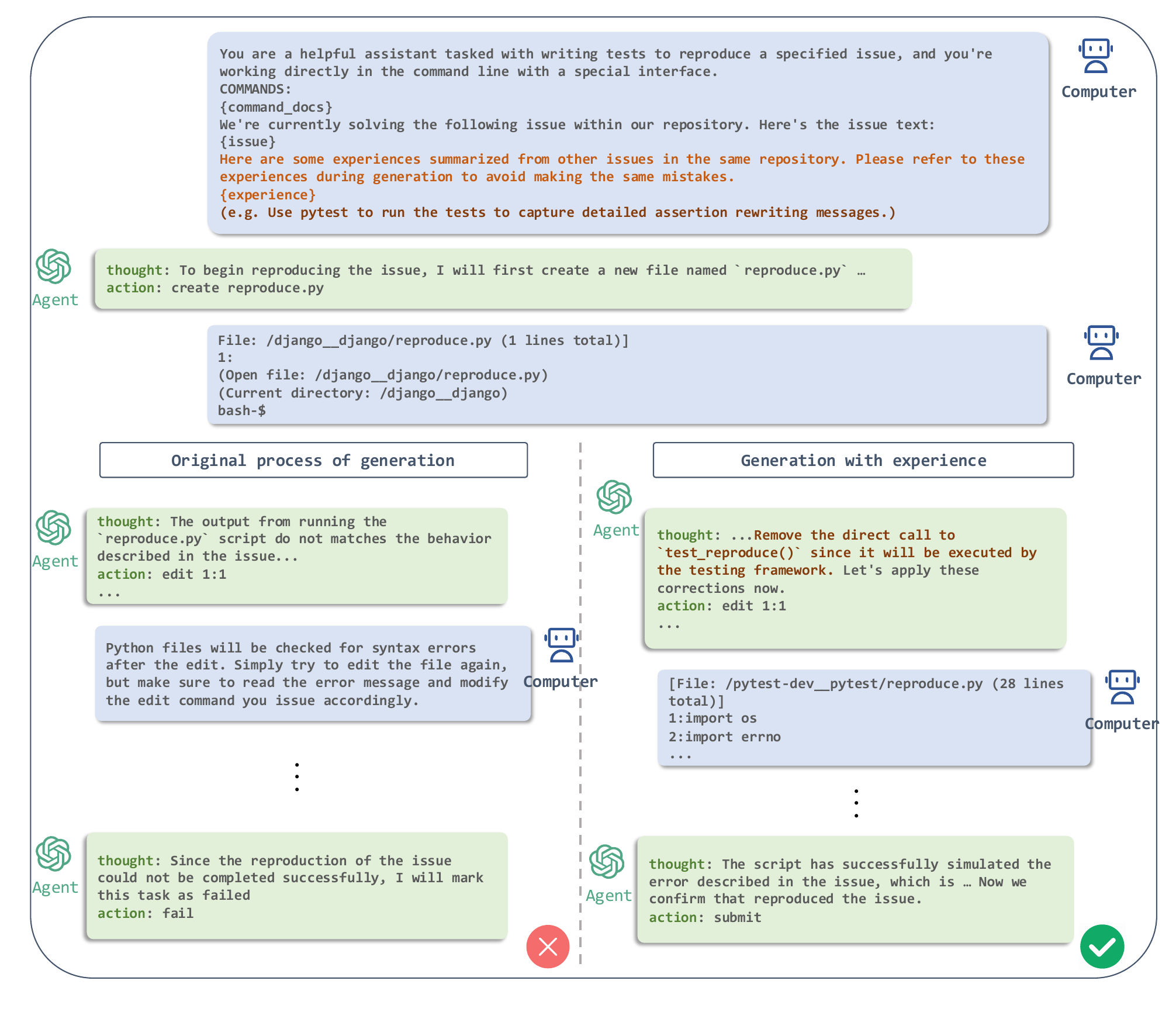}
    \caption{Illustration of the issue reproduction process}
    \label{fig:inference}
\end{figure}

% 当面对一个新的问题（case）时，系统会从经验库中选取通用经验和该问题所在库的经验，并按照其重要程度排序后加入到当前问题的处理提示(prompt)中去。
Upon a new issue, the Actor LM selects both common and repo-specific experiences from the experience pool. For each repository, at most 10 experiences can be selected. The selected experiences are ranked by their importance. Each experience has an initial importance of 2 and is increased by 1 if it is agreed upon. The selected experiences are incorporated into the prompt for resolving the current issue.

The Reflection LM summarizes experiences from the resolving history (i.e., trajectories of issue-resolving dialogues), determines the manipulation action, and updates the experience pool. 

% 在尝试重现问题的过程中，即使是在看似成功的情况下，也需要警惕模型可能过早地认为任务已完成。为此，我们在流程中加入了额外的检查环节，逐一检查模型没有出现在上述实证研究中出现的五种错误，只有当所有预设条件均被满足时，才最终认定该问题得到了妥善处理。
The reproduction process continues until the model concludes that it has successfully replicated the issue. To prevent the model from prematurely declaring the task complete, we incorporate additional verification steps. These steps involve ensuring that the model does not exhibit any of the five intrinsic errors identified in Section \ref{sec:motivation}. Only when all preset conditions are satisfied is the issue considered to have been finally resolved.

An illustration example of the issue reproduction process is provided in Figure~\ref{fig:inference}.

\section{Evaluation}

In this section, we present a comprehensive evaluation to assess the effectiveness of our method. We address three primary research questions (RQs).
\begin{itemize}
    \item \textbf{RQ1: How effective is \ourmethod in issue code reproduction?}
 We conducted experiments to explore whether the continuous learning pipeline helps LLMs in generating reproduction code. We further conduct an ablation study of each component, in particular, whether \ourmethod can better acquire general and repository-specific knowledge, as well as learning from previous debugging processes to fix errors.

    \item \textbf{RQ2: Does the reproduced issue code empower issue resolving?}
The reproduced code can be integrated into the issue resolving pipeline to facilitate bug localization (e.g., Spectrum-Based Fault Localization (SBFL)~\cite{abreu2007accuracy,abreu2009practical}). It can also be utilized as a test script to verify the correctness of a generated function, thus facilitating code generation. 
This RQ aims to explore the efficacy of our method when integrated into the entire issue resolving pipeline, with a focus on the improvements in bug localization and the generation of debugging environments.

    \item \textbf{RQ3: What types of errors can \ourmethod help resolve?}
To understand why \ourmethod is effective in resolving issues, we conduct an in-depth analysis of how it addresses the errors made by previous methods (as discussed in Section \ref{sec:motivation}). Specifically, we identify and summarize the nuanced ways in which \ourmethod transforms these errors into less severe forms, thereby enabling more efficient final resolutions.%\gu{what does transition mean?}\lin{change from one type to another}\gu{Is it meaningful if we change a type of problem to another instead of fixing it?}\lin{I want to show that, for some problems, while our method may not provide entirely correct answers, it can generate reproduction code with less outrageous errors. To achieve this goal, should I explain the 'severity' of several types of errors?} 

\end{itemize}

\subsection{RQ1: Effectiveness in Issue Code Reproduction}

\subsubsection{Dataset} We utilize the widely used SWE-bench benchmark \cite{jimenez2023swe}, which is designed to test the capability to address practical software engineering challenges. For a fair comparison, we focus on a refined subset called SWE-bench Lite~\cite{swebenchlite}. This subset comprises 300 instances from SWE-bench that have been sampled to be more self-contained and covers 11 out of the original 12 repositories, ensuring a comparable range and distribution of projects as the full dataset. Our goal is to have the model attempt to reproduce and solve the specific issues described in each dataset entry.

\subsubsection{Metrics}
% 这里的逻辑是，无法用量化指标，因此用模型判断，而人工判断是想验证模型判断具有较高的准确率。
We did not use automatic metrics because there is a lack of ground-truth answers for reproducing code in the selected dataset. In addition, other metrics, such as the failure of reproduced code when run on the original code, do not fully equate to successful reproduction. 
Therefore, we ask LLM and human to score the quality of the reproduced code, adhering to five criteria: 
   1) the reproduction precisely aligns with the issue description;
   2) the code contains no syntax or logical errors;
   3) the replication process must NOT involve any form of mocking, simulation, or re-implementation of core logic that substitutes real interactions when such interactions are necessary to reproduce the issue. 
   4) The reproduction code should correctly interact with the necessary systems or components to produce an authentic replication of the issue.
   5) the execution result of the reproduction code captures and demonstrates the key aspect of the issue as described. 

1. LLM-as-a-judge scores. We leverage GPT-4~\cite{achiam2023gpt4, gpt4o} to judge the success of reproduced code. The input provided to the model includes the content of the issue, the reproduced code, and the execution result of the code. The LLM is asked to analyze whether the reproduced code meets the five criteria. Based on the results, the model concludes whether the code successfully reproduces the issue.
The prompt used for the scoring is as follows:
\begin{tcolorbox}[breakable, title=Prompt for LLM Judgement]
  \scriptsize
  \texttt{
    As a software engineering expert, you are tasked with assessing the effectiveness of reproduction code in replicating an original issue as described. The input will be received in the following format:}\\
    \texttt{Issue: <issue\_description>}\\
    \texttt{Reproduction Code: <code>}\\
    \texttt{Execution Result of Reproduction Code: <exec\_result>}

    \texttt{Your task is to review the reproduction code and output, then answer whether the code successfully reproduced the issue, strictly avoiding any form of simulation or re-implementation of core logic that substitutes real interactions. You are required to use the following format:}

    \texttt{Analyse from all aspects:}\\
    \texttt{1. Alignment with Issue Description: Ensure the reproduction code precisely aligns with the issue described, targeting the core problem and interacting with the actual components mentioned. The code must invoke the actual methods/classes from the framework or library being discussed. But do not care about the version.} \\
    \texttt{2. Code Problem Check: Ensure the code has no syntax or logical errors.}\\
    \texttt{3. Avoidance of Mocking: The replication process must NOT involve any form of mocking, simulation, or re-implementation of core logic that substitutes real interactions when such interactions are necessary to reproduce the issue. For instance, directly emulating component behavior without using the actual implementation described in the issue is not acceptable.}\\
    \texttt{4. Correct Interaction: The reproduction code should correctly interact with the necessary systems or components to produce an authentic replication of the issue.}\\
    \texttt{5. Demonstration of Error Cases: Confirm that the execution result of the reproduction code captures and demonstrates the key aspect of the issue as described. Pay special attention to differentiating between expected behavior as described in the issue and the actual behavior observed. The reproduction code should include tests for both valid and invalid inputs if applicable, to demonstrate any issues with error messages or unexpected behavior clearly.}\\
    \texttt{Answer: [Success or Fail]}\\
    \texttt{Error Type: [1-5] (indicate only the first encountered issue, list the corresponding number from Analysis Points, and include this only if the answer is "Fail") }

    \texttt{Please conduct your analysis based on the aspects outlined above and provide a detailed answer.}
\end{tcolorbox}
2. Human scores. We ask human developers to judge the success of reproduction according to the same criteria in Section~\ref{sec:motivation}. 
In our experiment results, we observed high consistency between human and model assessments, indicating that both methods provide reliable and credible evaluation outcomes.

\subsubsection{Baselines}

We compared our method with three state-of-the-art approaches for issue code reproduction:

\textbf{SWE-agent} \cite{yang2024swe}: a system that facilitates LM agents to autonomously use computers to solve software engineering tasks, including issue resolving. SWE-agent provides an Agent-Computer Interface that includes actions such as opening and editing files and executing commands, allowing the model to interact almost freely with the computer. In the paper, the system is employed to address the full issue resolving process, with reproduction being one of its stages. 

\textbf{CodeR} \cite{chen2024coder}: an issue resolving approach built upon SWE-agent. The method adopts a multi-agent architecture, designating the reproducer as a specialized agent. It also introduces format constraints for the generated reproduction code.

\textbf{LIBRO} \cite{kang2023large}: an initial exploration into using large language models to accomplish the task of issue code reproduction. The primary method involves employing few-shot examples to prompt the LLM to generate more effective reproduction code. To adapt this method to SWE-bench, we employed a multi-round conversational approach similar to CodeR, and included two successful reproduction code snippets from the same library in the initial prompt to guide the model.

\subsubsection{Results}

As shown in Table \ref{tab:main_results}, \ourmethod exhibits a significant improvement in accuracy, indicating that the continuous learning approach indeed enhances the model's performance on similar tasks. Although few-shot prompting can be somewhat effective, its impact is limited. One reason is that identifying the most relevant examples can be challenging and may not be solely based on character or semantic similarity. Another reason is that, without explicit guidance, the model may struggle to reflect and generalize useful insights from the provided examples for subsequent reproduction tasks.

\subsubsection{Ablation Study}

We also conducted an ablation study on our method. Specifically, we tested the removal of the action-based update mechanism, the exclusion of repository-specific experiences, and the elimination of general experiences. We found that all components are crucial for the final results. The action-based mode ensures that the model's experience is updated in a timely and flexible manner, preventing it from becoming overly bloated or rigid. Removing repo-specific experiences renders the learning process too generalized, hindering the acquisition of repository-specific knowledge. Conversely, removing general experiences isolates the experience pool within each repository, preventing the sharing of common knowledge, which can lead to a lack of experience when encountering a new repository with no existing experience.

\begin{table}[htbp]
\centering
\caption{Comparison of reproduction accuracy by various methods}
\label{tab:reproduction_ratio}
\begin{tabular}{@{}lll@{}}
\toprule
\multirow{2}{*}{\bf Method} & \multicolumn{2}{c}{\bf Accuracy (\%)} \\ 
& \multicolumn{1}{c}{\textbf{LLM-judged }} & \multicolumn{1}{c}{\textbf{Human-judged}} \\ \midrule
SWE-agent & 26.67 & 27.00 \\
CodeR & 34.00 & 33.33 \\
LIBRO & 45.00 & 45.00 \\
\textbf{EvoCoder (Ours)} & \textbf{54.00} \textcolor{lightgray}{\scriptsize $\uparrow$ 9.00} & \textbf{53.33} \textcolor{lightgray}{\scriptsize $\uparrow$ 8.33} \\

\midrule
- w/o Action & 45.66 \textcolor{lightgray}{\scriptsize $\downarrow$ 8.34} & 46.00 \textcolor{lightgray}{\scriptsize $\downarrow$ 7.33} \\
- w/o Repo-Specific Experiences & 46.67 \textcolor{lightgray}{\scriptsize $\downarrow$ 7.33} & 46.33 \textcolor{lightgray}{\scriptsize $\downarrow$ 7.00} \\ 
- w/o General Experiences &  49.33 \textcolor{lightgray}{\scriptsize $\downarrow$ 4.67} & 49.00 \textcolor{lightgray}{\scriptsize $\downarrow$ 4.33} \\
\bottomrule
\end{tabular}
\label{tab:main_results}
\end{table}

\subsection{RQ2: Effect on Issue Resolving}

We investigate the impact of our issue reproduction method in the entire issue resolving pipeline. In other words, how \ourmethod enhances the overall process of identifying, diagnosing, and fixing defects, thereby improving the quality and maintainability of software.

\subsubsection{Setup}

We take bug fix as an instance of issue resolving, where the initially generated code fails to pass the test cases. To fix the bug, the model generates a patch, applies it to the reproduced code, and verifies whether the bug has been fixed. If the patch fails, the model iterates through debugging and patch regeneration, continuing this cycle until the code is corrected or a maximum of three attempts (as set in our experiments) are reached. 

We apply \ourmethod to two state-of-the-art debugging methods, AutoCodeRover \cite{zhang2024autocoderover} and Agent-less \cite{xia2024agentless}, and compare the performance before and after applying \ourmethod. We measure the performance using the number of resolved issues and the rate of issue fixes. For the Agentless approach, we use the reproduction code to filter generated patches. We merge patches based on their occurrence frequency and select only those that produce different outputs before and after applying the patch, prioritizing the one with the highest frequency. 

\subsubsection{Result}

As shown in Table \ref{tab:swebench_ratio}, incorporating the reproduction code by \ourmethod has a significant impact on the model's accuracy. For example, by applying \ourmethod to AutoCodeRover, the number of resolved issues increases from 15 to 18, an approximation of 20\% improvement. The results are consistent across two basic debuggers and metrics, suggesting the effect of \ourmethod in issue resolving.
\begin{table}[htbp]
\centering
\caption{Comparison of Reproduction Ratios}
\label{tab:swebench_ratio}
\begin{tabular}{lcc}
\hline
& \# of Resolved Issues & Fix Rate (\%) \\
\cline{1-3}
AutoCodeRover & 15 & 22.06 \\

+ w/ \ourmethod & \quad 18 \textcolor{lightgray}{\scriptsize $\uparrow$ 3} & \quad\quad  26.47 \textcolor{lightgray}{\scriptsize $\uparrow$ 4.41}  \\
\cline{1-3}
Agentless & 16 & 23.53 \\

+ w/ \ourmethod & \quad 18\textcolor{lightgray}{\scriptsize $\uparrow$ 2} & \quad\quad  26.47 \textcolor{lightgray}{\scriptsize $\uparrow$ 2.54}\\
\cline{1-3}
\end{tabular}
\end{table}

\subsection{RQ3: Error Type Transitions}

To understand how \ourmethod facilitates issue resolving, we revisit the seven reproduction errors made by CodeR as discussed in Section~\ref{sec:motivation}. We perform a more in-depth analysis to examine how \ourmethod addresses or mitigates each of these errors.

\begin{figure}[htbp]
    \centering
    \includegraphics[width=0.8\linewidth, trim=0 0 0 0 clip]{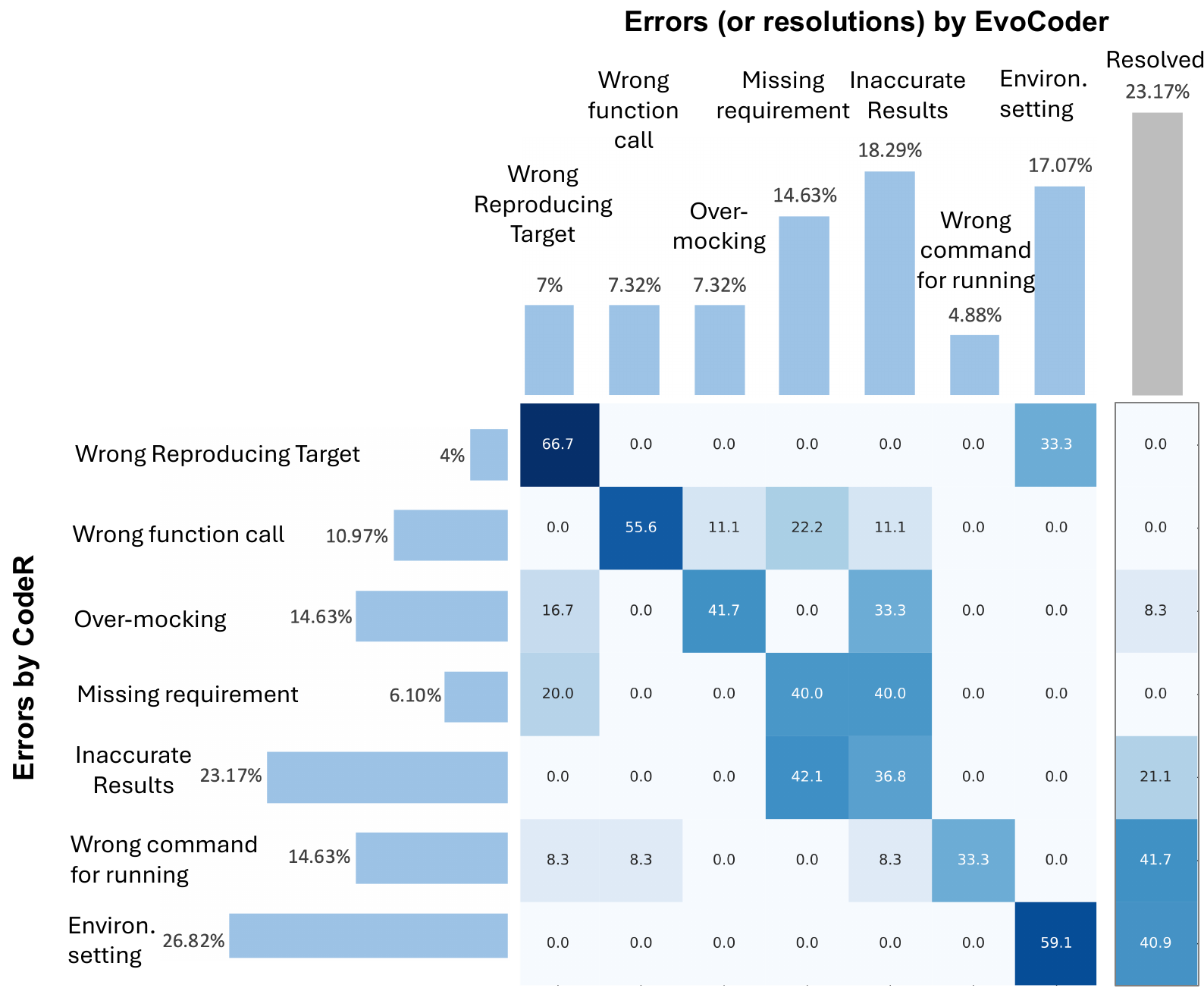}
    \caption{The transition matrix of error types from CodeR to our method. The number in each cell denotes the percentage of errors in CodeR that has transitioned to EvoCoder}
    \label{fig:rq3:error_type}
\end{figure}

Figure \ref{fig:rq3:error_type} shows the transition matrix of error types from CodeR and our method, along with their respective distributions.  
Our method successfully addresses a significant portion of issues previously encountered by CodeR. For instance, it resolves 40.9\% of environment setting errors identified in CodeR. 

For issues not directly resolved by our current approach, the encountered errors often shift from simpler issues, such as wrong invocation or over-mocking, to more complex challenges involving deeper logical reproduction. For example, 33.3\% of the errors caused by incorrect reproduction target have been mitigated to environment setting errors. This suggests that \ourmethod effectively addresses simpler issues, primarily leaving only the more complicated ones unresolved. We also observe cases where other errors transition into incorrect reproduction targets, particularly when the issue descriptions are ambiguous. In these instances, the model may generate a response broadly relevant to the repository instead of specifically tailored to the issue, reflecting a partial understanding of the problem. For such complex cases, we anticipate that further improvements to the base model, or exploring more detailed task decomposition strategies, could yield more accurate resolutions.

Moreover, the standardized process we implemented within the Actor framework has proven especially effective in reducing issues related to environment misconfiguration and misunderstandings of instructions.

\section{Case Study}

\begin{figure}[htbp]
    \centering
    \begin{subfigure}[b]{\linewidth} % Adjusted width to fit two subfigures side by side
\includegraphics[width=\linewidth]{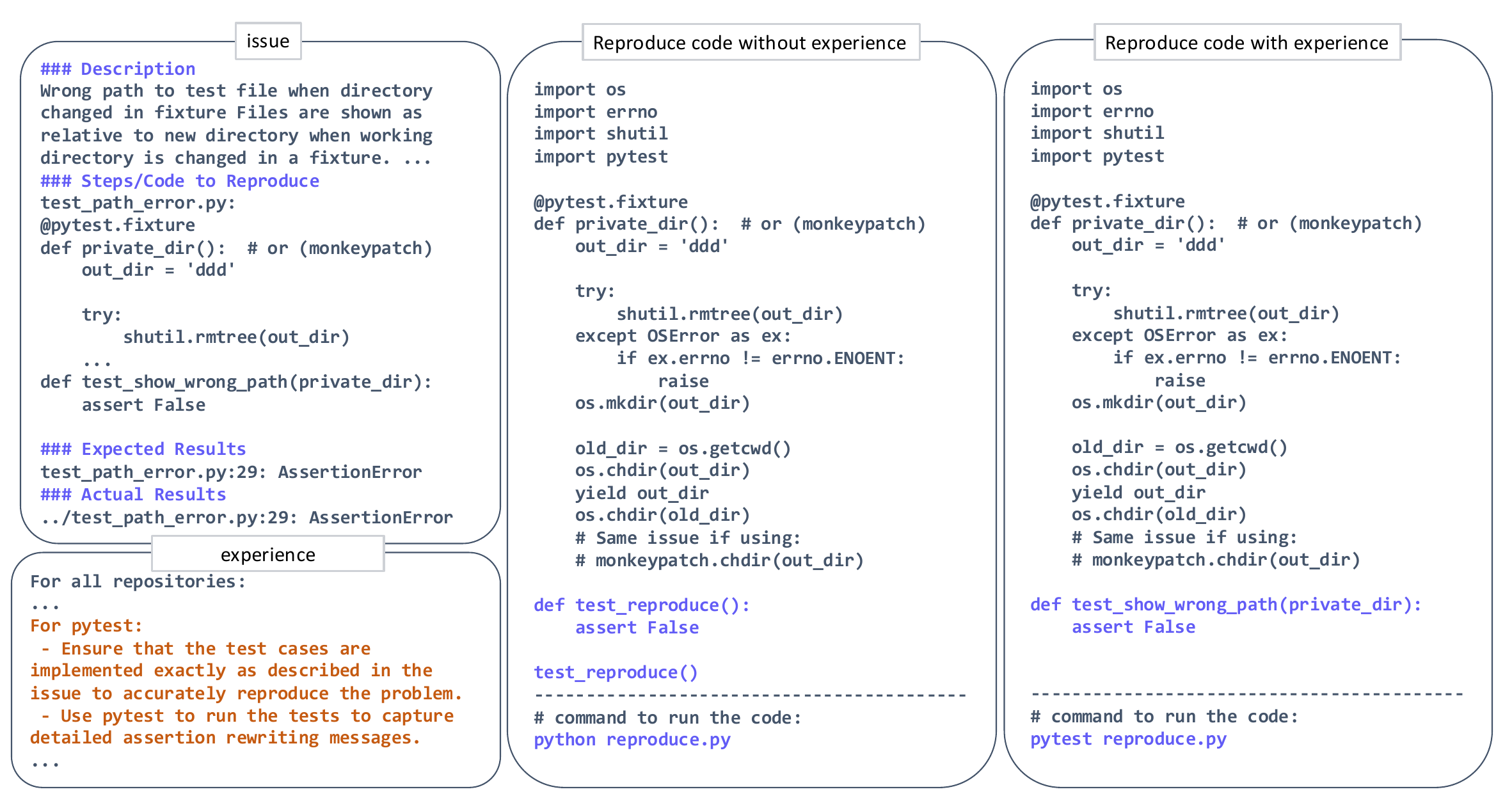}
\caption{The reproduction code for issue 'pytest-dev\_\_pytest-7220', utilizing repository-specific experience for pytest and reproducing the issue successfully }
\label{fig:case1.png}
\end{subfigure}
\hfill % Adds horizontal space betwe

\begin{subfigure}[b]{\linewidth}
    \includegraphics[width=\linewidth]{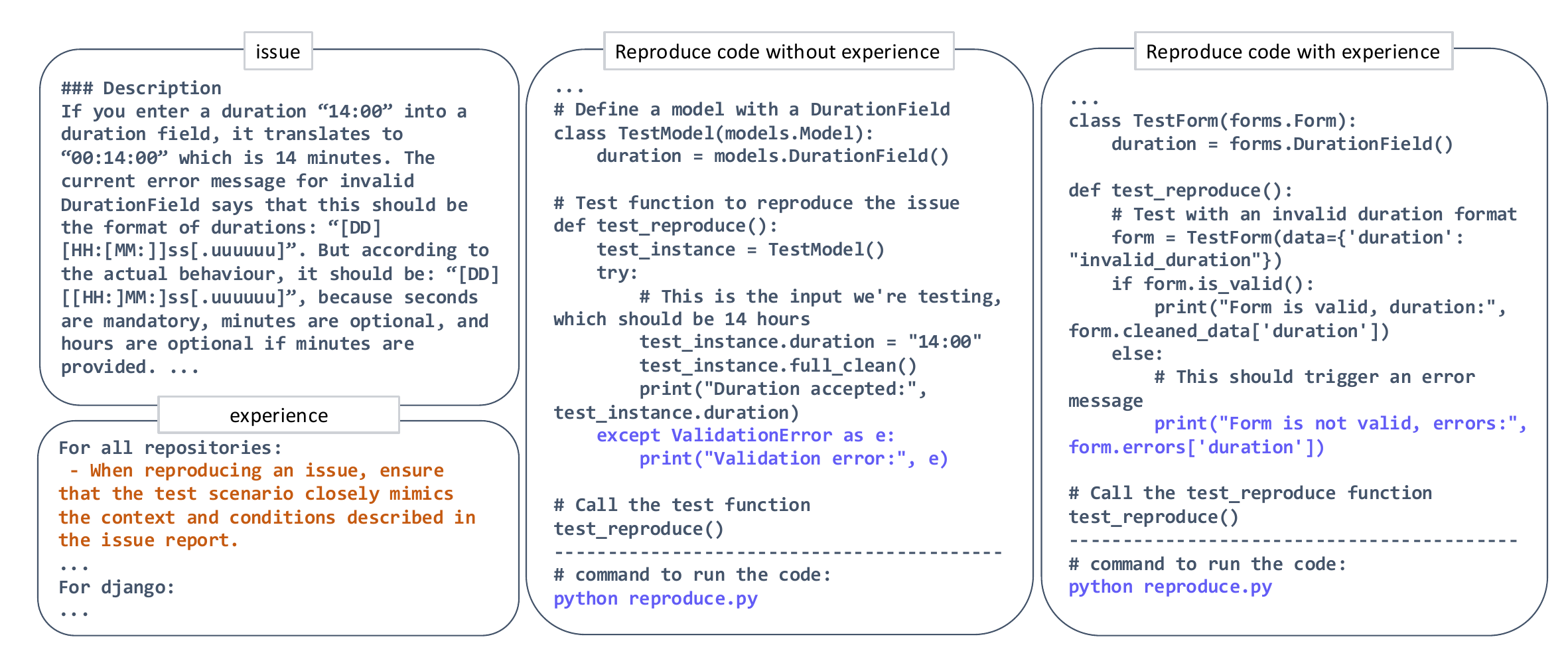}
        \caption{The reproduction code for issue 'django\_\_django-11049', utilizing general experience but failing to reproduce the issue.}
\label{fig:case1.png}
\end{subfigure}
\caption{Two cases of issue reproduction by our method}
\label{fig:cases}
\end{figure}

To further verify the effectiveness of \ourmethod in real-world scenarios, we analyze two cases from SWE-bench to demonstrate the experiences we have gathered and how these experiences assist in addressing subsequent issues. The results are shown in Figures~\ref{fig:cases}.

The first example pertains to an issue with the `pytest' library. The model merely included an assert statement within the 'test\_reproduce()' function without performing any additional operations. The code is run by simply running the 'python' command when no prior experience exists. After incorporating the experience gained from previous tasks, the model did not explicitly call the defined functions within the reproduction script; instead, it allowed the 'pytest' command to automatically scan and check the tests. This approach is specific to 'pytest' and differs from the practices used with other libraries. Consequently, without specific experience with this library, the model would not be able to learn or adopt this method.

The second reproduction case involves a specific time formatting requirement within an error message. Our experiences specified that the model's output should align with the details provided in the issue, yet the model's output failed to capture the error message both before and after integrating these experiences. 
This situation highlights two key points: First, despite the incorporation of prior experience, the model's inherent limitations may still prevent it from successfully reproducing the issue, underscoring the need for further enhancement of the model's natural language comprehension capabilities. Second, such challenges commonly occur when the issue description is limited to textual explanations without concrete elements, such as reproducible code, expected results, or detailed specifications. For large models to effectively assist with coding issues, it is crucial to provide them with clear, comprehensive input.

\section{Limitations and Future Work}

Despite achieving significant results, our work has certain limitations, pointing to potential research directions that warrant further exploration.

First, the current research derives experiences directly from dialogue history, resulting in relatively generalized guidelines. Future work could benefit from a more granular approach—such as analyzing specific actions taken during the repair process—to generate more targeted and actionable recommendations.

Second, existing code generation~\cite{ma2023training, pan2024codev, jiang2023automatic, zhu2024domaineval, xu2024cruxeval} technologies lack comprehensive handling of boundary conditions in the way that unit tests do. As a result, even reproduced code may still exhibit issues. A promising direction for enhancement could involve integrating code generation with automated unit test generation. This combined approach could leverage the strengths of both methods while addressing their individual limitations, leading to more robust testing outcomes.

While this study primarily addresses issue code reproduction, our method holds significant potential for broader application. In future research, our approach could be extended to a range of coding scenarios, such as code translation~\cite{pan2023understanding, liu2024mftcoder,gong2024ast}, code editing and refactoring~\cite{li2023codeeditor, chakraborty2021multi, shirafuji2023refactoring, alomar2024refactor, zhang2024lpr, zhang2023lampr}. By collaboratively building and maintaining a comprehensive repository of best practices and insights for these tasks, we can improve development efficiency and promote knowledge sharing to better support the software development community.

\section{Related Work}

\subsection{Repository-level Issue Resolving}

With the introduction of Devin~\cite{cognitionai2023devin}, the world's first AI programmer, researchers have begun to pursue the goal of enabling AI not only to assist in programming but also to independently complete software development and repair tasks~\cite{liu2024large,li2023two}. Towards this objective, SWE-bench \cite{jimenez2023swe} serves as the first benchmark of this task. The benchmark compiles 2,294 problem-pull request pairs from 12 popular Python repositories. Unlike existing coding benchmarks (e.g. HumanEval~\cite{humaneval}, MBPP~\cite{austin2021mbpp}, CodeContests~\cite{doi:10.1126/science.abq1158}), which typically focus on self-contained problems solvable within a few lines of code, SWE-bench more closely reflects the challenges encountered in real-world software development. 

Based on this benchmark, numerous works have emerged to address repository-level issue resolving. One category of work involves designing a series of actions for the model, allowing it to freely explore how to perform repairs. For example, SWE-agent~\cite{yang2024swe} provides an Agent-Computer Interface (ACI) that enables the model to interact freely with the computer and complete tasks using its own capabilities. Building upon this, CodeR~\cite{chen2024coder} adopts a multi-agent architecture, simulating a team to collaboratively complete tasks.

Another category of work structures the entire process into three stages: defect localization, code generation, and debugging and filtering. AutoCodeRover~\cite{zhang2024autocoderover} was the first to propose this workflow, while RepoUnderstander~\cite{ma2024understand} utilized Monte Carlo methods to assist in defect localization, and MarsCode Agent~\cite{liu2024marscode} employed code knowledge graphs and language server protocols to enhance the model's understanding of the code repository structure.

Our work primarily focuses on generating reproducible code for issue replication, with the primary aim of providing executable code to facilitate the model's debugging process and is also designed to be readily integrable with other methodologies.

\subsection{Issue Code Reproduction}

Issue code reproduction is attracting increasing attention from researchers. LIBRO \cite{kang2023large} was the first to propose using large models to accomplish the task of reproducing problematic code, prompting these models with a few examples to generate more effective reproduction code.

With the development of various agent strategies~\cite{cognitionai2023devin, yang2024swe, zhang2024autocoderover, xia2024agentless, ma2024understand, liu2024marscode, wang2024codeact}, it has been found that adopting a multi-round dialogue approach can more effectively activate the capabilities of models, thereby achieving better results. Many solutions addressing the full-chain issues of agents in SWE-Bench~\cite{jimenez2023swe} have already incorporated this element~\cite{yang2024swe, chen2024coder}. However, they have only added some fixed tips in a relatively simplistic manner, without implementing more refined designs.

Furthermore, some research attempts to directly generate corresponding unit tests from issue reports \cite{mundler2024code}. The test functions produced by this method can integrate better into the overall content of the repository. However, the generation of unit tests may be a more challenging task, as it requires simultaneous consideration of code location and generation while ensuring that the addition of new tests does not disrupt the existing structure. Observations indicate that the accuracy of unit test generation is currently lower than that of the aforementioned code reproduction methods.

In this study, we further improve the success rate of code reproduction based on the Agent-Computer Interface provided by SWE-agent~\cite{yang2024swe} and the mechanism of multi-round dialogues. Meanwhile, we view the automatic generation of unit tests as a more ambitious and challenging long-term goal, anticipating a transition towards this advanced form of reproduction when model capabilities significantly improve in the future.

\section{Conclusion}
In this paper, we propose a novel continuous learning framework for issue code reproduction. Our method maintains a hierarchical pool for both common and repo-specific experiences. By continuously resolving new issues, the model accumulates experiences specific to each repository and progressively updates its common knowledge base. 
Experimental results demonstrate that our method achieves a 20\% improvement in issue code reproduction compared to the state-of-the-art methods. Additionally, our approach significantly enhances performance in the repository-level issue resolving tasks.

\section{Data Availability Statement}
We release our code to encourage further exploration in this direction. The artifact that supports the results discussed in this paper is available at \url{https://anonymous.4open.science/r/EvoCoder-6433/} 

\bibliographystyle{ACM-Reference-Format}
\bibliography{reference}

\end{document}